\documentclass[sigconf]{acmart}

\AtBeginDocument{%
  }

\copyrightyear{2025}
\acmYear{2025}
\setcopyright{rightsretained}
\acmConference[CHI EA '25]{Extended Abstracts of the CHI Conference on Human Factors in Computing Systems}{April 26-May 1, 2025}{Yokohama, Japan}
\acmBooktitle{Extended Abstracts of the CHI Conference on Human Factors in Computing Systems (CHI EA '25), April 26-May 1, 2025, Yokohama, Japan}\acmDOI{10.1145/3706599.3720162}
\acmISBN{979-8-4007-1395-8/2025/04}

\usepackage{textcomp}

\begin{document}
\newcommand{\Sys}{\textsc{ConversAR}}

\title[\Sys{}: Exploring Embodied LLM-Powered Group Conversations in AR for Second Language Learners]{\Sys{}: Exploring Embodied LLM-Powered Group Conversations in Augmented Reality for Second Language Learners}

\author{Jad Bendarkawi}
\authornote{indicates equal contribution by authors}
\email{jadb@princeton.edu}
\affiliation{%
  \institution{Princeton University}
  \city{Princeton}
  \state{New Jersey}
  \country{USA}
}
\author{Ashley Ponce}
\authornotemark[1]
\email{ap36@princeton.edu}
\affiliation{%
  \institution{Princeton University}
  \city{Princeton}
  \state{New Jersey}
  \country{USA}
}
\author{Sean Chidozie Mata}
\authornotemark[1]
\email{sm5607@princeton.edu}
\affiliation{%
  \institution{Princeton University}
  \city{Princeton}
  \state{New Jersey}
  \country{USA}
}
\author{Aminah Aliu}
\email{aa1237@princeton.edu}
\affiliation{%
  \institution{Princeton University}
  \city{Princeton}
  \state{New Jersey}
  \country{USA}
}
\author{Yuhan Liu}
\email{yuhanl@princeton.edu}
\affiliation{%
  \institution{Princeton University}
  \city{Princeton}
  \state{New Jersey}
  \country{USA}
}
\author{Lei Zhang}
\email{raynez@princeton.edu }
\affiliation{%
  \institution{Princeton University}
  \city{Princeton}
  \state{New Jersey}
  \country{USA}
}
\author{Amna Liaqat}
\email{al0910@princeton.edu}
\affiliation{%
  \institution{Princeton University}
  \city{Princeton}
  \state{New Jersey}
  \country{USA}
}
\author{Varun Nagaraj Rao}
\email{varunrao@princeton.edu}
\affiliation{%
  \institution{Princeton University}
  \city{Princeton}
  \state{New Jersey}
  \country{USA}
}
\author{Andrés Monroy-Hernández}
\email{andresmh@princeton.edu}
\affiliation{%
  \institution{Princeton University}
  \city{Princeton}
  \state{New Jersey}
  \country{USA}
}
\renewcommand{\shortauthors}{Bendarkawi, Ponce, Mata, Aliu, et al.}


\begin{abstract}
Group conversations are valuable for second language (L2) learners as they provide opportunities to practice listening and speaking, exercise complex turn-taking skills, and experience group social dynamics in a target language. However, most existing Augmented Reality (AR)-based conversational learning tools focus on dyadic interactions rather than group dialogues. Although research has shown that AR can help reduce speaking anxiety and create a comfortable space for practicing speaking skills in dyadic scenarios, especially with Large Language Model (LLM)-based conversational agents, the potential for group language practice using these technologies remains largely unexplored. We introduce \Sys, a gpt-4o powered AR application, that enables L2 learners to practice contextualized group conversations. Our system features two embodied LLM agents with vision-based scene understanding and live captions. In a system evaluation with 10 participants, users reported reduced speaking anxiety and increased learner autonomy compared to perceptions of in-person practice methods with other learners.

\end{abstract}




\begin{teaserfigure}
   \includegraphics[width=1\textwidth]{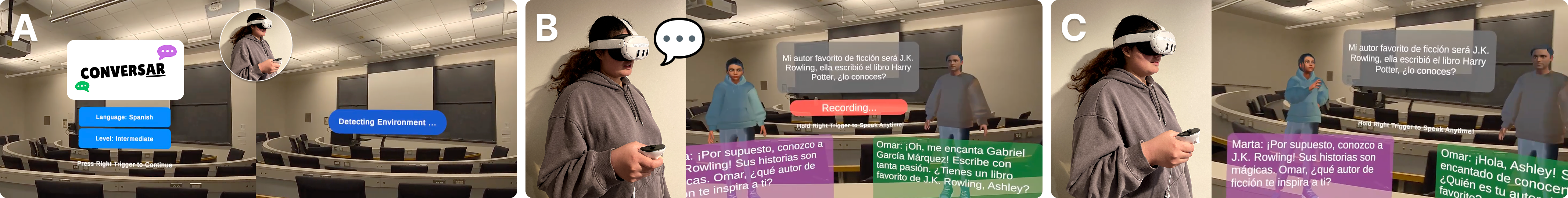}
   \caption{Demonstration of the \Sys{} system. (A) Title screen where the user selects the language (Spanish) and proficiency level (intermediate) then observes the system detecting the surrounding environment. (B) The user holds the right controller trigger to speak into the system. (C) One of the LLM agents responds to the user with live captions appearing underneath them.}
   \Description[Screenshots of a user interacting with the system]{The figure shows a user interacting with an Augmented Reality system for group conversations among second-language learners (L2). Panel A displays the initial interface where the user selects their language (Spanish) and proficiency level (Intermediate) while the system detects the surrounding environment. Panels B and C illustrates the user wearing a VR/AR headset and participating in a group conversation with virtual avatars, whose dialogue is displayed in Spanish, creating an immersive language-learning experience.}
   \label{}
\end{teaserfigure}

\maketitle

\section{Introduction}
Group conversation poses a considerable challenge for second-language (L2) learners, as it demands advanced communicative competence while simultaneously managing the anxiety associated with speaking in group settings \cite{young_creating_1991, macintyre_subtle_1994}. In fact, a majority of L2 learners report difficulty participating in group discussions due to anxiety and linguistic limitations \cite{horwitz_foreign_1986}. While recent advances in Augmented Reality (AR) and Large Language Models (LLMs) have created promising opportunities for conversational learning like narrative-based learning games \cite{cheng_scientific_2024, zhao_language_2024}, tools for practicing contextualized language use \cite{lee_visionary_2023, hollingworth_fluencyar_2023}, and immersive interactions with embodied agents \cite{sousa_calepso_cardlearner_2022, pan_ellma-t_2024, liu_developing_2024}, existing systems mainly focus on dyadic (one-on-one) interactions. These implementations, while valuable, fail to capture the complexity of group conversations that characterize many practical communication scenarios. Group conversations require greater mastery over turn-taking, responding to different conversational styles, and in-context use of a wide variety of speech functions (e.g., suggestions, agreement or disagreement, explanations, etc.)
\cite{stroud_second_2017, gan_interaction_2010}. 

We present \Sys{}, a novel system that leverages AR and embodied LLM agents to simulate group conversations for L2 learners. \Sys{}{} is a Meta Quest 3 app that situates two virtual, gpt-4o powered, humanoid agents within a user's physical environment. The agents can understand the user's environment through object detection, take turns dynamically, and respond in a target language with natural voices and captions. We evaluated the system with 10 university students in the US with an intermediate or higher level of Spanish who used the system for 10 minutes and answered a survey and set of semi-structured interview questions. We found that participants experienced reduced speaking anxiety, increased learner autonomy, and greater active participation compared to perceptions of in-person practice methods. We also identified challenges such as limited emotional investment and competing visuals. 

\section{Related Work}

\subsection{Group Conversations for L2 Learners}
Practicing group conversations is a key aspect of mastering second language acquisition (SLA). Group conversation uniquely offers the opportunity for L2 learners to simultaneously practice listening and oral skills \cite{jarquin_tapia_importance_2022}. Learners can observe and adopt strategies to keep conversations flowing, such as turn-taking, interrupting, and active listening \cite{ernst_talking_1994}. Additionally, when learners control group conversation topics they are more likely to use diverse strategies to overcome communication challenges, enhancing their L2 skills \cite{ernst_talking_1994}. Likewise, several studies have found that both the quantity and quality of language production improved when language learners were in small groups as opposed to direct teacher instruction \cite{ellis_language_2012}. However, group conversations can be difficult for L2 learners due to anxiety, especially in classroom settings \cite{tavares_role_2016, jarquin_tapia_importance_2022}, which can negatively impact learning outcomes \cite{felicity_speaking_2018}. 

\subsection{Leveraging Extended Reality and Large Language Models for
SLA}
\subsubsection{Extended Reality for SLA}
 Research has demonstrated improved motivation levels \cite{chen_effects_2022} and vocabulary acquisition \cite{zhi_extended_2023} in Extended Reality (XR) applications for SLA. Specifically, AR and Virtual Reality (VR) have a positive impact on SLA through contextualized language learning opportunities \cite{lee_it_2024, liu_beyond_2023, schorr_foreign_2024}. While language immersion can accelerate SLA, especially as it relates to sociolinguistic competence \cite{chacon-beltran_acquisition_2010}, traveling abroad to target language countries or finding a tutor might be inaccessible, expensive, or unapproachable. For the anxious learner, AR and VR provide a safe environment for facilitating more authentic communication in L2 \cite{lee_it_2024, schorr_foreign_2024} while being more immersive than other form factors (mobile, web, etc). Also, the situated nature of these virtual environments immerses learners in contexts that mimic daily language use, enhancing their ability to retain and apply the language in relevant practical scenarios \cite{brown_situated_1989, lee_it_2024, schorr_foreign_2024, mills_potentials_2022}.
 
\subsubsection{LLM-based Conversational Agents for SLA}
Extensive research indicates that conversational agents that interact with students using natural language and human-like personalities can enhance engagement, resulting in greater learning outcomes \cite{johnson_pedagogical_2018, sonlu_effects_2024, kim_exploring_2024, abbasi_measuring_2014}. In the context of SLA, LLM-powered chatbots have demonstrated the ability to generate coherent and contextually relevant text for learning scenarios leading to greater retention of vocabulary, incidental vocabulary learning  \cite{zhang_impact_2024, cong_demystifying_2025, cong_ai_2024}, as well as increasing learners' willingness to communicate \cite{ayedoun_conversational_2015}. However, text-based tools alone lack the body language and audiovisual cues critical to human conversation, reducing learner interest \cite{huang_revisiting_2022}. As such, researchers have shown the value of embodied LLM agents with expressive avatars in improving engagement and learning outcomes \cite{sonlu_effects_2024, pan_ellma-t_2024, lefkowitz_second_2009}.

\subsubsection{A Novel Approach}
While prior systems have integrated a singular embodied LLM agent in VR \cite{pan_ellma-t_2024} or combinations of a text-based agent with scene understanding in AR \cite{lee_visionary_2023, hollingworth_fluencyar_2023}, the implementation of multiple embodied agents in AR for SLA remains largely unexplored. 
Although there are larger-scale explorations like CILLIE \cite{divekar_foreign_2022}, an XR system that uses a human-scale 360\textdegree \space panoramic screen for multi-agent group conversations, implementations like these are not portable nor accessible for frequent use given the chosen medium. Our work contributes to the existing literature by proposing the first system that leverages AR with multiple embodied LLM agents to enable real-time L2 group conversation practice.

\begin{figure*}
    \centering\includegraphics[width=0.95\linewidth]{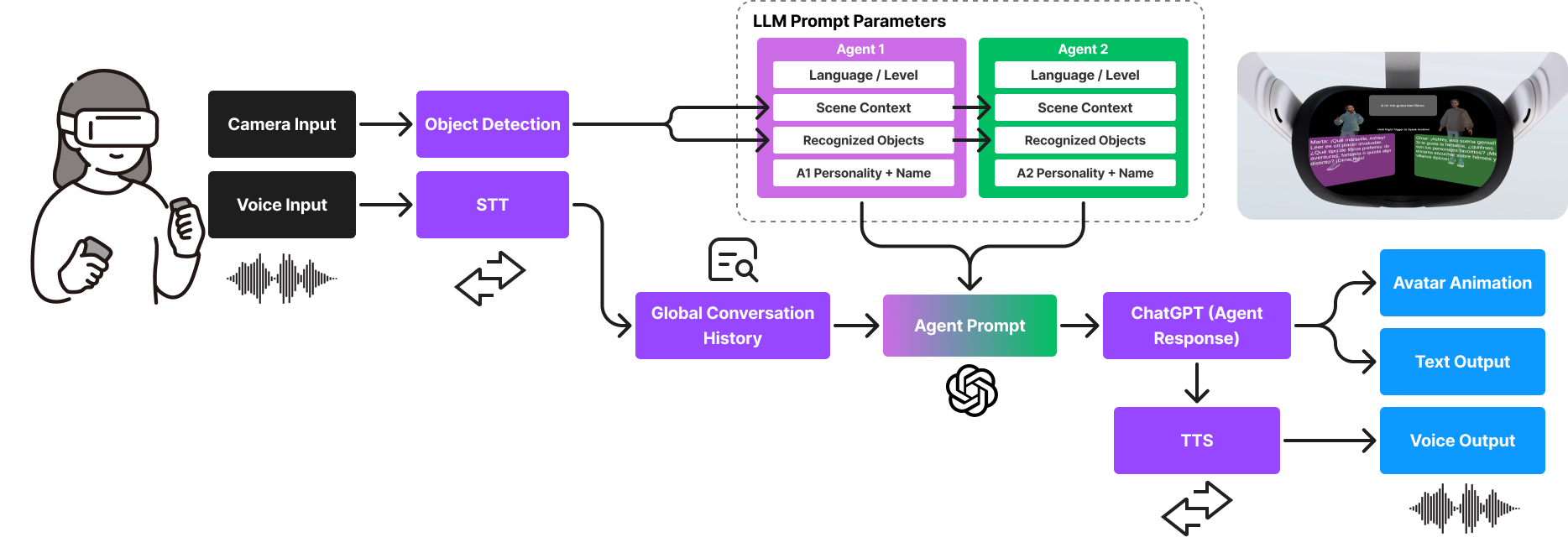}
    \Description[A flowchart of the system]{
The figure depicts a system design for a language learning application using virtual reality and AI. On the left, a person wearing a VR headset and holding controllers which represents the user. Two inputs are shown: ``Camera Input'' leading to ``Object Detection'' and ``Voice Input'' leading to ``STT'' (Speech-to-Text). Outputs from these modules connect to a ``Global Conversation History'' module, which feeds into the ``Agent Prompt.'' Two agents, labeled Agent 1 and Agent 2, are configured with attributes such as language level, scene context, recognized objects, and personality/name. The parameters contribute to the ``Agent Prompt'' processed by an LLM. Responses are then used to generate avatar animations, text output, and TTS (Text-to-Speech) voice output, displayed in the VR headset interface with agent dialogues.}
    \caption[System architecture of \Sys{}. The architecture highlights core components and data flow within the system.]{System architecture of \Sys{}. The architecture highlights core components and data flow within the system.\footnotemark[1]
    }

    \label{fig:gameflow}
\end{figure*} 

\section{The \Sys{} System}
\Sys{}{} is an AR, Meta Quest 3 application developed in Unity that enables L2 learners to engage in group conversations with two embodied LLM agents in a target language. The system incorporates advanced voice recognition and response capabilities through integration with the ChatGPT Audio Model (whisper-1) and Chat Model (gpt-4o). Before a conversation begins, the system takes a snapshot of the user's environment, performing object detection to enable agents to engage the user in contextually relevant dialogue. The agents take turns conversing with the user and each other, with the user being able to interject, answer, or ask questions to either of them at any point. 

 We specifically chose an AR headset form factor as it provides the benefits of immersion and engagement observed in VR conversational systems \cite{pan_ellma-t_2024, liu_beyond_2023} while uniquely enabling us to interweave context from the users physical world into the experience through scene understanding \cite{lee_visionary_2023, hollingworth_fluencyar_2023}. Where fully VR systems would be limited by pre-determined 3D rendered environments, by using an AR approach, users can engage in conversations that reflect their surroundings for more relevant L2 practice across diverse contexts.

We iteratively developed and refined the system by conducting a pilot study with two adult university undergraduate students enrolled in an intermediate Spanish course. Each pilot tester had a 10-minute conversation using the system and was briefly interviewed, providing insights for improving conversation flow, agent behavior, and interaction design. 

\subsection{Object Detection \& Contextualized Learning}
Recent works exploring the joint application of LLMs and AR for SLA have utilized object detection to allow L2 learners to engage in contextualized discussions about their physical surroundings \cite{lee_visionary_2023, hollingworth_fluencyar_2023}. Studies show that this contextualized learning approach more effectively promotes language acquisition and retention than classroom-based methods \cite{ellis_empirical_1997} and boosts learner engagement and motivation \cite{norris_effectiveness_2000}. Similarly, we adopt object detection for scene understanding into our system to enable contextually relevant conversations. The Meta Quest 3 restricts our ability to access or store camera feedback from the device, thus we opted for a Wizard of Oz approach in this iteration of \Sys{}. 

Before participant use, one of the researchers hardcodes a list of objects and the scene context onto the device based on the location of use. Users observe a ``Detecting Environment...'' loading indicator before the conversation begins to create the illusion that the system has detected their immediate surroundings. Although this process involves human intervention, it mimics the responsiveness and accuracy expected from an automated system and can create more consistent environmental context parameters across user studies.

\subsection{Agent Embodiment \& Persona}
Drawing inspiration from existing embodied conversational learning systems \cite{pan_ellma-t_2024, divekar_foreign_2022}, users engage with two humanoid agents that are animated and have distinct natural voices from OpenAI's audio library. Studies indicate that agents with natural-sounding voices reduce cognitive load, allowing learners to focus more on language processing rather than decoding speech \cite{schwieter_cognitive_2022}, which is especially important in the L2 context. Additionally, we utilize a cartoon rendering style as prior work proves this aesthetic makes agents appear as more agreeable than uncanny photorealistic styles \cite{sonlu_effects_2024, cassell_emotion_2000, zibrek_does_2014}. Agent personality also plays a critical role in learner comfort and information retention \cite{sonlu_effects_2024}, thus we prompt each agent to have unique yet generally playful and helpful personalities.

\subsection{Conversation Workflow}
\subsubsection{Prompt System}
To minimize agent response latency, we use a zero-shot prompting approach \cite{sahoo_systematic_2024}. A stand-alone prompt containing all task specifications and constraints guides the LLM towards the desired task. The LLM used in our system is not fine-tuned. This approach uses less resources than methods that require extensive example cases or multi-step reasoning \cite{sahoo_systematic_2024}. 
The prompt is parameterized to take in the target language and level, scene context, and global conversation history so far (Fig. \ref{fig:gameflow}). We refined the prompt to emphasize engaging both agents and the user equally, preventing one-sided interactions while facilitating ample speaking opportunities. We also include guardrails in the prompt to keep the conversation aligned with the scene's context while allowing some freedom to explore related topics, preventing significant tangents.

\begin{figure*} 
    \centering
    \includegraphics[width=\linewidth]{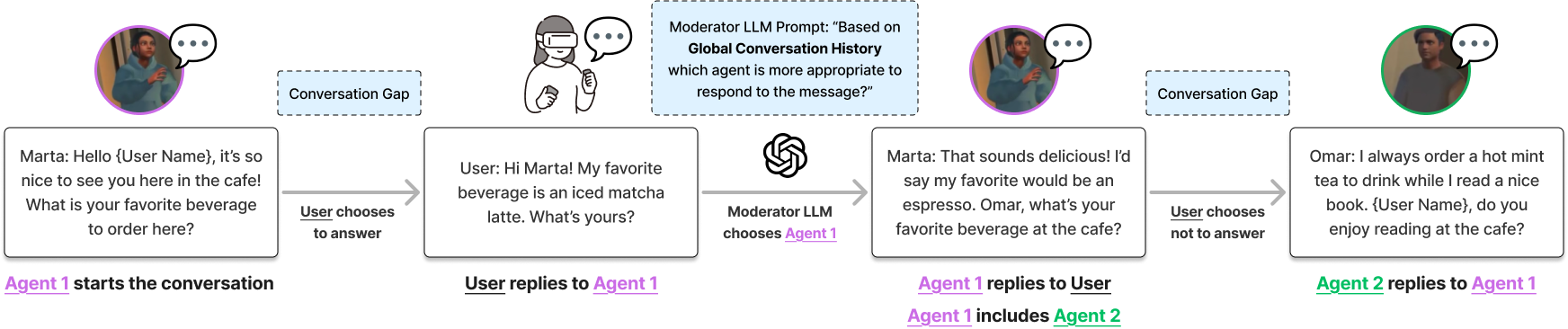}
    \Description[A flowchart of an example conversation]{The figure illustrates an example conversation flow between a user and two agents in a virtual environment. The interaction begins with "Agent 1" (Marta), represented by an avatar with a speech bubble, greeting the user: "Hello {User Name}, it's so nice to see you here in the cafe! What is your favorite beverage to order here?" The user, depicted wearing a VR headset, responds: "Hi Marta! My favorite beverage is an iced matcha latte. What's yours?" The system, moderated by an LLM, evaluates the "Global Conversation History" to determine which agent should respond next. The LLM selects Agent 1, who replies, "That sounds delicious! I'd say my favorite would be an espresso. Omar, what's your favorite beverage at the cafe?" Agent 2 (Omar), represented by a second avatar, responds: "I always order a hot mint tea to drink while I read a nice book. {User Name}, do you enjoy reading at the cafe?".}
    \caption[Example conversation flow between a user and the two agents in English, situated in a café. After the user speaks, the "\texttt{Moderator LLM}"  selects Agent 1 as the appropriate respondent. Agent 1 responds, then attempts to include Agent 2 in the conversation.]{Example conversation flow between a user and the two agents in English, situated in a café. After the user speaks, the "\texttt{Moderator LLM}"  selects Agent 1 as the appropriate respondent. Agent 1 responds, then attempts to include Agent 2 in the conversation.\footnotemark[1]
    }
    \label{fig:convo}
\end{figure*} 

\footnotetext[1]{\textit{Woman Wearing A Vr Head SVG} Vector by Soco St, licensed under \href{https://creativecommons.org/licenses/by/4.0/}{CC BY 4.0}. Available via \href{https://soco-st.com/?ref=svgrepo.com}{SVG Repo}. Recolored.}

\subsubsection{STT-TTS Pipeline} 
We utilize the OpenAI Audio API's speech-to-text (STT) and text-to-speech (TTS) modules to handle all voice-based inputs and outputs. Users interact with the agents by holding the trigger button on the right Meta Quest Touch controller and speaking in the target language. Although we considered voice-based triggers, our pilot studies suggested that the button interaction was intuitive and felt familiar to using a walkie-talkie. This approach also supports mid-speech pausing, addressing a limitation of fully voice-based conversational systems that struggle to distinguish whether the user intends to stop speaking or is pausing to think \cite{pan_ellma-t_2024}. When the user lets go of the trigger, their speech is saved as an mp3 file that is passed through the "whisper-1" model and is finally returned as text output to prompt agent responses. Agent responses are verbalized via the Audio API and displayed in live caption windows underneath each agent, following accessible caption placement recommendations from prior AR research \cite{jain_exploring_2018}.

The STT module exhibited high accuracy but occasionally auto-corrected user speech inputs in ambiguous cases (e.g., unclear pronunciation, filler words, incorrect tense usage) to maintain grammatical consistency, which pilot testers found helpful for error recognition. The TTS model also performed well, with pronunciation evaluated as natural and correct by a native Spanish-speaking researcher. However, there was an observable delay when transitioning from STT to TTS. While this delay was imperceptible during agent-to-agent interactions where the next utterance is preemptively generated and stored, STT user inputs consistently resulted in a 1-2 second pause before TTS agent responses. This delay scaled proportionally with longer, more complex speech inputs, however, did not impede the experience across pilot and study sessions.

\subsubsection{Conversation Turn Management} A gpt-4o powered \texttt{"Moderator LLM"} transitions the conversation between \texttt{agent-speaking} and \texttt{user-speaking} states. During the \texttt{agent-speaking} state, the conversation oscillates between the two agents indefinitely. The agents direct responses to each other and the user, but ultimately the user decides when to reply or interject. For example, one agent may ask the user a direct question, but if they do not respond during a defined conversational gap, the other agent will respond instead. This approach was designed to prepare users for conversation dynamics in which a delayed response or lack of contribution leads to other interlocutors continuing the discussion. This is a common reality for L2 learners, especially when engaging in group conversations dominated by more proficient or native speakers \cite{tavares_role_2016}. Through pilot studies, we determined an optimal gap of 3 seconds between each agent-generated response. Users can transition to the \texttt{user-speaking} state at any time by holding the right controller trigger, initiating STT. After the user responds, the \texttt{"Moderator LLM"} analyzes the conversation history to select the most appropriate responding agent (Fig. \ref{fig:convo}). This setup allows users to continue the activity indefinitely.

Turn taking and management of a mixed-initiative conversation in conversational agents is a complex and ongoing research problem \cite{novick_conversational_2018}, especially because human interaction relies on multi-modal cues such as prosodic, semantic, and syntactic modulation alongside physical behaviors like gaze and gestures \cite{ehret_whos_2023}. The advantage of our current approach is that it is flexible and simple to implement, however, future work may integrate perception of these multi-modal cues to better emulate human conversational flow.

\section{Evaluation}
To evaluate our system, we recruited 10 adult university undergraduate students (4 male, 6 female, aged 18-23) who have taken or are currently taking an intermediate or higher level Spanish course (Appendix \ref{appendixParicipants}). We recruited participants via emails and word of mouth. Each participant was compensated \$15 USD after an hour-long study. We conducted the study in a university library, with the agents prompted to discuss favorite books and literary genres. First, we showed participants a 5-minute walk-through video, then each participant engaged in a 10-minute group conversation using the system. We recorded each conversation. Next, participants completed a survey with 7-point Likert scale questions about  system usability, their perceptions of the system and task, and their perceptions of in-person L2 group conversations (Appendix \ref{appendixSurvey}). Finally, we conducted a  \(\sim \)30-minute semi-structured interview about the overall experience, perception of the activity, perception of system features, engagement with system features, and open feedback (Appendix \ref{appendixInterview}). 

Following a prior study in L2 task design \cite{lambert_learner-generated_2017}, we measured conversational engagement metrics across behavioral, cognitive, and social dimensions. Behavioral engagement refers to the degree of active involvement in the task, cognitive engagement refers to sustained attention and mental effort, and social engagement refers to a degree of reciprocity and mutual involvement during the task \cite{lambert_learner-generated_2017}. One researcher, who is a native Spanish speaker, collected these metrics from the conversation recordings to contextualize the interview responses (Appendix \ref{AppendixCodedRecording}). In addition, the researcher gave each participant a score based on the Advanced Placement (AP) interpersonal speaking rubric as a proxy for conversational depth \cite{ap_spanish_2018} (Appendix \ref{appendixParicipants}). Our university's Institutional Review Board reviewed and approved our study.

\section{Results}
The results of our study indicate that \Sys{} \space provided several benefits to L2 learners, including reduced speaking anxiety, increased learner autonomy, and greater willingness for linguistic risk-taking. 
Participants indicated they experienced similar levels of engagement but less speaking anxiety using \Sys{} in comparison to in-person practice methods (Appendix \ref{appendixAnxietyGraph}). Participants found the system easy to use (M = 6.0, SD = 0.82), and would use the system again in the future (M = 6.0, SD = 0.94). In addition, participants felt that the overall experience was immersive and simulated in-person interactions well, P1 noted: \textit{``I found [it] no different than participating in a trio for conversations in class. I think that there was hardly any discernible difference between if I had actual human Spanish speakers in front of me.''} The following sections elaborate on these findings, categorized into the key benefits and challenges of the system.

\subsection{Key Benefits of LLM-powered Group Conversations in AR}

\subsubsection{Reduced speaking anxiety} 

Participants reported feeling more comfortable speaking and expressing themselves while using \Sys{} compared to in-person learning methods. The majority of participants found that the system allowed them to remain relaxed and speak more freely, without the fear of being judged. As P3 observed, \textit{``[The] aspect that it's not a real person kind of allows for more authenticity. You're more open to making a mistake. You're not as fearful ... I could just express myself however and not have to think too hard about certain things.”} Participants further appreciated how the system allowed them to control their level of engagement to manage dynamics that would otherwise have been anxiety-inducing. P7 noted, \textit{``I like that if I didn't have a response to this specific question, I didn't have to answer it, and the other people would just continue talking, which made it a little bit less stressful.”} Additionally, P6 valued the flexibility to ask for clarifications to the agents, noting how it \textit{``was really helpful, not to just move on, but to acknowledge that I don't know [something] … Whereas in the Spanish class, they don't always stop for you”}. 

\subsubsection{Increased learner autonomy} 
Participants reported an increased sense of autonomy and confidence while engaging with the system. Here, we refer to autonomy as the capacity for learners to tailor or self-direct the learning experience to suit their needs and interests. Participants particularly appreciated how the interaction with the LLM agents allowed for conversations that dynamically aligned with familiar topics and interests, making it easier and more engaging to participate. P3 noted that the conversation felt personalized, enabling them to navigate the dialogue with greater agency, \textit{``I was able to speak a lot [and] use words that I already know and they kind of drive the conversation because I was familiar with the topic so it made for a better and fruitful conversation''}. This adaptability gave users the flexibility to direct conversations towards topics of their choice, which they described as empowering, \textit{``I had choice [over] my conversation and how I wanted to steer the conversation''} (P3). 

\subsubsection{Increased risk-taking} 
Participants expressed feeling empowered to take linguistic risks for several reasons. One key factor was their ability to mimic the language usage and new vocabulary demonstrated by the conversational agents. For example, P2 highlighted, \textit{``I would have used a more basic word, but I saw the version [the agents] used and ... just [tried] to see if I can on-the-go use [it]”}. Similarly, P3 noted,  \textit{``It made me think of ... other vocabulary I could use that I wasn't really thinking too much about”}. Additionally, the lack of emotional risk in interacting with the simulated environment encouraged participants to experiment freely with their language skills.
P5 remarked that \textit{``there's not really the [same] emotional payoff of conversing with a human and taking risks … It might encourage you to be even more risky because there's no social backlash”}. 

\subsubsection{Agents foster active participation} 
Participants consistently perceived the agents as encouraging and genuinely invested participants in the conversation. By using positive affirmations and displaying distinct personalities with quirks and personal opinions, the agents created a supportive and engaging environment. This perception motivated learners to remain involved and take greater initiative in dialogue. P3 noted the agents \textit{``would respond [with] \textit{‘Oh how amazing!’} and stuff like that to reassure you [throughout] the conversation [which] makes you want to continue the conversation. That could sometimes be lacking in real life conversations”}. 

\subsection{Challenges of LLM-powered Group Conversations in AR}
\subsubsection{Emotional investment \& social realism}
Participants were aware that the agents were not real, which removed the incentive to foster genuine connections or take emotional risks. Underscoring this lack of social realism, P5 engaged in a meta-conversation with the agents, attempting to explore their capacity for sentience, noting \textit{``I started trying to get the AI to talk about their emotions ...  [but] they did seem confined to this idea of books [so I wanted to see] if you could break out of that.”} P5 asked questions like \textit{``How do you experience love?''} and \textit{``What are your greatest insecurities?''}, to which the agents gave surface-level responses and a follow-up steering back to the prompted discussion topic. P5 further commented on the restrictive nature of the prompt guardrails mentioning that, \textit{``[the agents] would always bring it back to books and literature, which is not how people work.”} This balance between allowing agents to deviate from the prompt to bring a sense of realistic spontaneity to the interaction versus remaining grounded in the scene context posed a challenge. When the agents adhered too strictly to the discussion topic, participants reported the conversation felt more constrained and socially unnatural, however, when the agents deviated too far from the user's scene context, it risked disconnection from environmental relevance, which could undermine the value of contextual learning.

\subsubsection{Visual elements compete for user attention} 
Participants noted that, while they were very focused during the task, their high attention towards captions sometimes detracted from their attention to the embodied agents. Users often found themselves prioritizing the captions in order to follow the discussion more effectively. P1 mentioned, \textit{``I noticed [the agents] a few times … but I was more so focused on their voices and the text box (live captions), so there was less contact with said avatars”}. Some participants also commented on the minimal expressivity of the agents, which possibly made them less visually salient compared to the captions. Based on prior research, we speculate that more expressive, non-anthropomorphic embodiments, such as cartoon animals \cite{ahmed_design_2021} with more varied and relevant non-verbal cues (i.e., nodding, facial expressions, eye gaze) could more effectively aid real-time comprehension \cite{davis_meta-analytic_2023, cassell_power_1999}. If users can gain more context from non-verbal expressions, they may not as readily default to the captions for understanding,
possibly promoting more balanced attention across the visual elements.

\section{Discussion}

Our findings highlight the value of group-based settings in fostering effective L2 conversational practice in ways that are distinct from dyadic experiences. The presence of multiple agents uniquely provided opportunities to observe interactions passively before engaging, which is inherently missing from one-on-one experiences \cite{ngah_exploring_2019}. Unlike dyadic conversations, where there is an expectation for more rapid and continuous speaking patterns \cite{pouw_timing_2022}, the group setting enabled participants to speak at their own discretion, fostering a sense of ease and controlled participation.

While we recognize the benefits of LLM agents in providing opportunities to learn and mimic new language use and vocabulary, we must also consider the implications of practicing with fully fluent speakers compared to similar-leveled peers. Prior research shows that lower proficiency L2 learners demonstrate increased effort to reach shared understanding and less reliance on their first language in task-based oral activities with higher-level interlocutors over similar-leveled peers \cite{tian_l2_2021}. On the other hand, users may miss the distinct affective and emotional benefits of similar-level peer interactions \cite{zabihi_proficiency_2022}, such as greater instances of laughter and expressions of enjoyment, which may influence  learner motivation \cite{liu_beyond_2023}. 
It's also important to consider that group settings may introduce a higher cognitive load for some L2 learners. The added complexity of group interactions may overwhelm less confident learners, potentially hindering their ability to actively participate or process the conversation effectively \cite{ahmadi_interactional_2019, stroud_second_2017}. This suggests that dyadic systems may be more effective for such learners until a higher level of L2 proficiency and confidence is built.

\section{Limitations and Future Work}

The sample size of our study was relatively small (n = 10), limiting the generalizability of our findings. Although we recruited participants from Spanish courses of similar proficiency levels (Appendix \ref{appendixParicipants}), variability in speaking skills made it difficult to isolate system effects. We also acknowledge that study-related performance pressure may have influenced speaking performance. 

Our comparison against in-person methods relied solely on self-report measures, thus, future studies should more rigorously evaluate differences between in-person group conversation practice methods and interventions like \Sys. 
Additionally, unlike the single-session format used in this preliminary work, conducting a longitudinal study would enable an analysis of the system's long-term effects on learning outcomes. Also, future studies should directly compare multi-agent versus dyadic conversation and examine the impact of varying the number of agents. Investigating system performance across diverse physical contexts (eg., café, office, etc.) could yield insights into how contextualized learning shapes  engagement and learning outcomes. Further studies may also explore how groups of agents can be adapted to simulate combinations of pedagogical roles (e.g. a fellow student and teacher) or more complex multi-party role-play scenarios. 

Finally, participants provided valuable suggestions for additional features, such as gamification, toggleable live captions, and error corrections. Users also recommended more expressive virtual agents, proposing a tighter integration of LLM outputs with embodied behaviors (e.g., facial expressions) which could improve social and emotional engagement.

\section{Ethics \& Accessibility}
While we acknowledge the opportunities the Quest 3 provides for accessing novel learning tools, we recognize that its cost is inaccessible to many demographics of learners. Furthermore, our project was designed by Spanish language learners and native speakers in our research group, which may have unintentionally prioritized the needs and preferences of learners with similar linguistic, educational, and cultural backgrounds. Additionally, human-like pedagogical agents, while innovative, raise ethical concerns about job displacement in language instruction and the loss of nuanced, empathetic and cultural interactions that come from engaging with real people. Currently, \Sys{} relies on hand-trigger interactions with the Meta Quest Touch controllers, which poses accessibility barriers for individuals who cannot use tactile interfaces. Future iterations of the system should incorporate alternative input methods, such as voice commands or gesture recognition to ensure our system is inclusive of learners with different physical abilities.

\section{Conclusion}
Our research explores the potential of LLMs and AR to provide enhanced and accessible group conversation opportunities for L2 learners. We successfully developed and evaluated an AR and LLM-powered headset system, equipped with scene understanding, embodied agents, and live speech-to-text and text-to-speech capabilities. Our findings suggest that participants perceived our system as highly beneficial for L2 learners looking to practice group speaking in their target language. Compared to in-person methods, the system was associated with reduced speaking anxiety and comparable engagement levels. Key insights from the study highlight that participants felt the system provided a safe and comfortable environment for L2 group conversation and increased learner autonomy. This comfort and agency are crucial for L2 learners, as they foster more open and active participation, which is intimately linked to learning outcomes. Despite the study's limited duration, users expressed confidence that continued use would lead to long-term improvement in their language skills. Looking ahead, we envision expanding the system to support more languages, integrate additional agents, and incorporate accessible hands-free controls. Future research should focus on improving the social and emotional interaction between the LLM agents and users, investigating long-term system effects on performance in L2 group conversations, and exploring areas where the system's user experience can be enhanced.

\section{Disclosure of the Usage of AI Tools}
We used ChatGPT (GPT-4o model) to facilitate the writing of this manuscript. This includes code assistance for  turning raw data into charts and graphs, correcting grammar and spelling mistakes, and polishing existing writings with prompts like ``Find me a synonym of X'' and ``Shorten this sentence without changing its content''.

\bibliographystyle{ACM-Reference-Format}
\bibliography{bibver6}
\balance

\clearpage

\nobalance
\appendix
\section{Appendix}
\subsection{Interview Guide} \label{appendixInterview}

\textit{** This is a 30 minute semi-structured interview. We may ask follow up questions or skip questions based on the participant's engagement and quality of responses. **}
    \subsection*{Warm up and participant background}
    \begin{enumerate}
        \item What is your name and age?
        \item Could you describe your Spanish language level?
        \item How long have you been speaking Spanish?
        \item If you are comfortable sharing, when you are practicing Spanish speaking, what kinds of things do you find difficult / what challenges do you encounter?
        \item How would you describe your comfort level when participating in group conversations in Spanish?
        \item Could you describe your access to opportunities to participate in Spanish group conversations?
    \end{enumerate}

    \subsection*{Overall perception}
    \begin{enumerate}
        \item What was your experience like overall?
        \item How would you describe the usability of the system? Was it comfortable/uncomfortable to use? Did you find anything particularly unclear about using the system?
        \item What benefits / value do you see to using the system?
    \end{enumerate}

    \subsection*{Perception of activity}
    \begin{enumerate}
        \item In general, how did you feel while participating in the task? What emotions, if any, did you experience while participating in the conversation?
        \item How would you describe your comfort level interacting with the LLM agents?
        \item How would you describe your level of confidence during the activity?
        \item How would you describe your speaking anxiety level during the activity?
        \item Did you experience any moments where the conversation felt awkward or unnatural? If so, can you describe them?
        \item Did you feel like you had enough time to interject?
        \item Did you feel like the conversation matched the context of your environment?
        \item Were there any moments you felt like you lost focus during the activity?
        \item How would you describe what you learned from the task, if anything?
    \end{enumerate}

    \subsection*{Perception of features}
    \begin{enumerate}
        \item How did you feel about the avatar's body language during the conversation?
        \item How do you think your experience would change if the avatars weren't there and you were interacting with just text/audio?
        \item How did you feel about seeing the captions?
        \item What did you think about the AR / immersion aspect of the system?
        \item What did you think about the object detection / contextual aspect of the system?
        \item How well did the conversations flow with the agent?
        \item How would you describe the authenticity or naturalness of your conversations with the agents?
        \item How would you describe the feeling of authenticity when using \Sys{}?
    \end{enumerate}

    \subsection*{Engagement with features}
    \begin{enumerate}
        \item How would you describe your level of engagement in the conversation compared to the in-person group conversation?
        \item How would you describe your level of confidence speaking in the conversation compared to the in-person group conversation?
        \item How would you describe your level of speaking anxiety in the conversation compared to the in-person group conversation? If you felt this at all?
        \item How would you describe your level of participation in the conversation compared to the in-person group conversation?
        \item How would you describe your motivation level to continue improving your Spanish after the \Sys{} conversation compared to the in-person group conversation?
        \item How would you describe the effectiveness of \Sys{} for practicing group conversations?
        \item Did you feel like the system encouraged you to improve your language skills, such as trying new vocabulary or sentence structures?
        \item Did you feel like the system encouraged you to take risks during the conversation?
        \item If the system had no technical issues and was widely available, would you use it for practicing group conversations? If so, how frequently?
    \end{enumerate}

    \subsection*{Open feedback}
    \begin{enumerate}
        \item Is there anything you would want to change about the experience?
        \item Any other comments you'd like to share?
    \end{enumerate}

\subsection{Post-activity Survey} \label{appendixSurvey}

\subsection*{System Workload Demand, Ease-of-use, Future Adoption}

Answer the following questions reflecting on your experience completing the AR group conversation task using \Sys{}.

\begin{enumerate}
    \item How mentally demanding was the task?
    \begin{itemize}
        \item Very Low \dotfill 1 \dotfill 2 \dotfill 3 \dotfill 4 \dotfill 5 \dotfill 6 \dotfill 7 \dotfill Very High
    \end{itemize}
    \item How physically demanding was the task?
    \begin{itemize}
        \item Very Low \dotfill 1 \dotfill 2 \dotfill 3 \dotfill 4 \dotfill 5 \dotfill 6 \dotfill 7 \dotfill Very High
    \end{itemize}
    \item How hurried or rushed was the pace of the task?
    \begin{itemize}
        \item Very Low \dotfill 1 \dotfill 2 \dotfill 3 \dotfill 4 \dotfill 5 \dotfill 6 \dotfill 7 \dotfill Very High
    \end{itemize}
    \item How successful were you in accomplishing what you were asked to do?
    \begin{itemize}
        \item Perfect \dotfill 1 \dotfill 2 \dotfill 3 \dotfill 4 \dotfill 5 \dotfill 6 \dotfill 7 \dotfill Failure
    \end{itemize}
    \item How hard did you have to work to accomplish your level of performance?
    \begin{itemize}
        \item Very Low \dotfill 1 \dotfill 2 \dotfill 3 \dotfill 4 \dotfill 5 \dotfill 6 \dotfill 7 \dotfill Very High
    \end{itemize}
    \item How insecure, discouraged, irritated, stressed, and annoyed were you?
    \begin{itemize}
        \item Very Low \dotfill 1 \dotfill 2 \dotfill 3 \dotfill 4 \dotfill 5 \dotfill 6 \dotfill 7 \dotfill Very High
    \end{itemize}
    \item How much do you agree with this statement: The system was easy to use?
    \begin{itemize}
        \item Strongly Disagree \dotfill 1 \dotfill 2 \dotfill 3 \dotfill 4 \dotfill 5 \dotfill 6 \dotfill 7 \dotfill Strongly Agree
    \end{itemize}
    \item Would you consider using \Sys{} again in the future?
    \begin{itemize}
        \item Definitely no \dotfill 1 \dotfill 2 \dotfill 3 \dotfill 4 \dotfill 5 \dotfill 6 \dotfill 7 \dotfill Definitely yes
    \end{itemize}
\end{enumerate}

\subsection*{\Sys{} Conversation Activity}

Questions related to the AR group conversation task you completed.

\begin{enumerate}
    \item How engaged did you feel while using \Sys{}?
    \begin{itemize}
        \item Barely / not engaged at all \dotfill 1 \dotfill 2 \dotfill 3 \dotfill 4 \dotfill 5 \dotfill 6 \dotfill 7 \dotfill Highly engaged for a long duration
    \end{itemize}
    \item How anxious did you feel when speaking during the \Sys{} group conversation?
    \begin{itemize}
        \item Not at all anxious \dotfill 1 \dotfill 2 \dotfill 3 \dotfill 4 \dotfill 5 \dotfill 6 \dotfill 7 \dotfill Extremely anxious
    \end{itemize}
    \item How confident did you feel when speaking during the \Sys{} group conversation?
    \begin{itemize}
        \item Not at all confident \dotfill 1 \dotfill 2 \dotfill 3 \dotfill 4 \dotfill 5 \dotfill 6 \dotfill 7 \dotfill Extremely confident
    \end{itemize}
    \item How would you rate your level of participation in the \Sys{} group conversation?
    \begin{itemize}
        \item Barely any / no participation \dotfill 1 \dotfill 2 \dotfill 3 \dotfill 4 \dotfill 5 \dotfill 6 \dotfill 7 \dotfill Extremely active participation
    \end{itemize}
    \item How motivated do you feel to practice and improve your language skills after using \Sys{}?
    \begin{itemize}
        \item Not at all motivated \dotfill 1 \dotfill 2 \dotfill 3 \dotfill 4 \dotfill 5 \dotfill 6 \dotfill 7 \dotfill Extremely motivated
    \end{itemize}
\end{enumerate}

\subsection*{Reflecting On In-person Methods}

Questions related to in-person methods for practicing group conversations. Here, \textbf{"in-person methods"} refers but is not limited to: in-class group conversations, Spanish language tables, in-person group tutoring, language immersion programs, etc.

\begin{enumerate}
    \item How engaged do you feel during in-person methods for practicing group conversations?
    \begin{itemize}
        \item Barely / not engaged at all \dotfill 1 \dotfill 2 \dotfill 3 \dotfill 4 \dotfill 5 \dotfill 6 \dotfill 7 \dotfill Highly engaged for a long duration
    \end{itemize}
    \item How anxious do you feel when speaking during in-person methods for practicing group conversations?
    \begin{itemize}
        \item Not at all anxious \dotfill 1 \dotfill 2 \dotfill 3 \dotfill 4 \dotfill 5 \dotfill 6 \dotfill 7 \dotfill Extremely anxious
    \end{itemize}
    \item How confident do you feel when speaking during in-person methods for practicing group conversations?
    \begin{itemize}
        \item Not at all confident \dotfill 1 \dotfill 2 \dotfill 3 \dotfill 4 \dotfill 5 \dotfill 6 \dotfill 7 \dotfill Extremely confident
    \end{itemize}
    \item How would you rate your level of participation during in-person methods for practicing group conversations?
    \begin{itemize}
        \item Barely any / no participation \dotfill 1 \dotfill 2 \dotfill 3 \dotfill 4 \dotfill 5 \dotfill 6 \dotfill 7 \dotfill Extremely active participation
    \end{itemize}
    \item How motivated do you feel to practice and improve your language skills after in-person methods for practicing group conversations?
    \begin{itemize}
        \item Not at all motivated \dotfill 1 \dotfill 2 \dotfill 3 \dotfill 4 \dotfill 5 \dotfill 6 \dotfill 7 \dotfill Extremely motivated
    \end{itemize}
\end{enumerate}
\newpage

\subsection{Speaking Anxiety \& Engagement Quantitative Results} \label{appendixAnxietyGraph}
\begin{figure}[h!]
    \centering
    \includegraphics[width=0.9\linewidth]{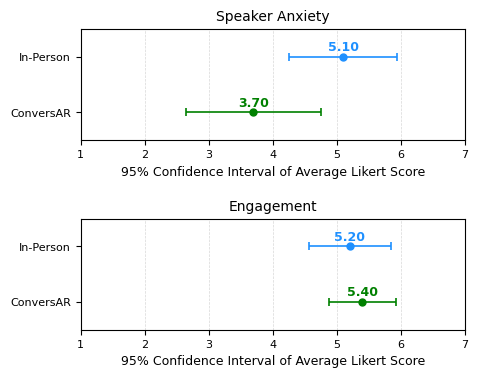}
    \caption{Graphs comparing user response on the perceived levels of speaking anxiety and engagement using our system versus in-person methods.}
     \Description[Graphs comparing using the system versus in-person in engagement and speaking anxiety]{Two horizontal bar graphs compare the 95\% confidence intervals of average Likert scores for speaker anxiety and engagement in two conditions: \Sys{} and In-Person. In the top graph, which represents speaker anxiety (where a lower score is better), the average score for \Sys{} is 3.70, with a narrower confidence interval, while the average score for In-Person is 5.10, with a wider confidence interval. This indicates that speaker anxiety is lower in the \Sys{} condition. In the bottom graph, which represents engagement (where a higher score is better), the average score for \Sys{} is 5.40, slightly higher than the score for In-Person, which is 5.20. The confidence intervals for both conditions overlap, suggesting similar levels of engagement.}
\end{figure}

\subsection{Cognitive, Behavioral, \& Social Engagement Across Participants} \label{AppendixCodedRecording}

One of the researchers who is a native Spanish speaker independently coded the recordings of participants' conversations using the system. The following metrics were tallied: behavioral engagement via number of turns taken, cognitive engagement via elaborative clauses and negotiations of meaning, and social engagement via number of backchannels (verbal cues to affirm active listening to the other interlocutors i.e \textit{"mm-hm"}, nodding, etc.) \cite{lambert_learner-generated_2017}. Behavioral engagement was higher than cognitive engagement across all participants besides P9. We recorded no instances of social engagement, demonstrating a gap in emulating the social reciprocity present in human L2 interactions.

\begin{figure}[h!]
    \centering
\includegraphics[width=\linewidth]{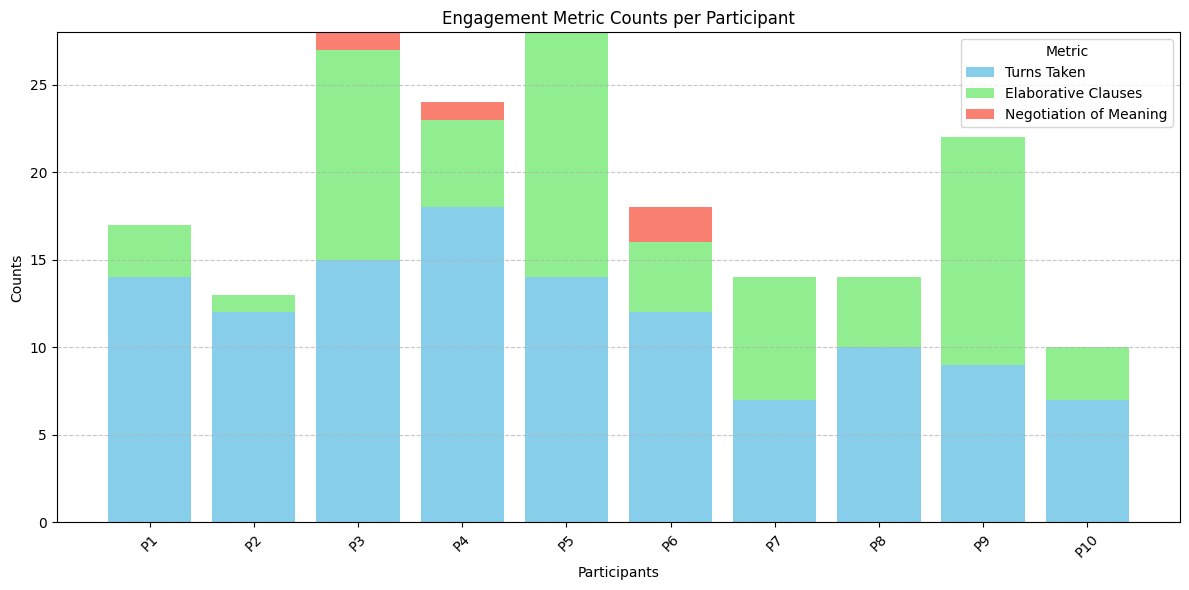}
    \caption{Engagement Breakdown per Participant. The number of turns taken, elaborative clauses and negotiations of meaning are plotted in blue, green and red respectively.}
   \Description[Engagement Metric Counts per Participant]{The figure is a stacked bar chart that displays the counts of three engagement metrics—turns taken, elaborative clauses, and negotiation of meaning for each participant (P1 to P10). Each participant's bar is divided into these metrics, illustrating variations in engagement across participants. Behavioral engagement (measured by turns taken) dominates for most participants, while cognitive engagement (elaborative clauses and negotiation of meaning) is lower. Participant P9 shows a unique distribution with a higher balance of cognitive engagement. No metric  tracking social engagement appears on the chart, highlighting a gap in the system's ability to simulate human-like social reciprocity in L2 interactions.}
\end{figure}

\onecolumn
\subsection{Participants Table \& Interpersonal Speaking AP Scores}
\label{appendixParicipants}
One of the researchers who is a native Spanish speaker independently listened to recordings of participants' conversations using the system and scored them on the AP Interpersonal Speaking Score scale as a proxy for conversational depth \cite{ap_spanish_2018}.

\begin{table*}[h!]
\centering
\begin{tabular}{ c c c c c c }
\hline
\textbf{participant} & 
\textbf{gender} & 
\textbf{age} & 
\textbf{Spanish class level} &
\textbf{L2 speaking anxiety (1-7)}  &
\textbf{conversational depth (1-5)}  \\
 \hline
P1  & F & 18 & advanced & 6 & 3\\ 
P2  & F & 18 & intermediate/advanced & 4 & 1\\ 
P3  & F & 19 & intermediate/advanced & 2 & 5\\ 
P4  & M & 21 & intermediate/advanced & 6 & 2\\ 
P5  & M & 23 & advanced & 5 & 5\\
P6  & M & 21 & intermediate/advanced & 7 & 4\\ 
P7  & F & 20 & advanced & 6 & 4\\ 
P8  & F & 22 & advanced & 5 & 4\\ 
P9  & F & 19 & intermediate/advanced & 5 & 5\\ 
P10 & M & 20 & intermediate/advanced & 5 & 2\\
\hline
Mean&& & & 5.1 & 3.5\\
\hline
\end{tabular}
\newline
\caption{Participants table.}
\end{table*}

\end{document}